\documentclass[useAMS,usenatbib]{mn2e}
     \usepackage{xcolor} \usepackage{ulem} \usepackage{changebar}

        \colorlet{ins}{blue} \colorlet{del}{red}
\usepackage[unicode]{hyperref}
\usepackage{graphicx,amssymb,amsmath}
\title[Turbulence Induced Spin-Orbit Misalignment]{The Turbulent Origin of Spin-Orbit Misalignment in Planetary Systems}
\author[Fielding et al.]{Drummond B. Fielding$^{1}$, Christopher F. McKee$^{1,2}$, Aristotle Socrates$^{3}$, \newauthor Andrew J. Cunningham$^{4}$, and Richard I. Klein$^{1,4}$\\
 $^1$Astronomy Department, University of California, Berkeley, CA 94720\\
 $^2$Physics Department, University of California, Berkeley, CA 94720\\
 $^3$Jackson, WY 83001\\
 $^4$Lawrence Livermore National Laboratory, Livermore, CA 94550}

\date{Released \today}

\begin{document}
\maketitle

\label{firstpage}

\begin{abstract}
The turbulent environment from which stars form may lead to misalignment between the stellar spin and the remnant protoplanetary disk. By using hydrodynamic and magnetohydrodynamic simulations, we demonstrate that a wide range of stellar obliquities may be produced as a by-product of forming a star within a turbulent environment. We present a simple semi-analytic model that reveals this connection between the turbulent motions and the orientation of a star and its disk. Our results are consistent with the observed obliquity distribution of hot Jupiters. Migration of misaligned hot Jupiters may, therefore, be due to tidal dissipation in the disk, rather than tidal dissipation of the star-planet interaction.

\end{abstract}

\section{Introduction}

The observed orbital properties of hot Jupiters provide powerful diagnostics of the formation and evolution of planetary systems. The observation that the orbits of some hot Jupiters are misaligned with the spins of their host stars \citep{Winn:2009a} has the potential of casting light on the formation of protoplanetary disks (e.g., \citealt{Bate:2010}), on magnetic interactions between these disks and their host young stellar objects (e.g., \citealt{Lai:2011}), on dynamical interactions between the hot Jupiters and other massive planets or a binary companion in the systems (e.g., \citealt{Nagasawa}), and on tidal interactions between the hot Jupiters and their host stars (e.g., \citealt{RogersLin2013}). An issue of particular importance is, how did the hot Jupiters get so close to their host stars? Under the plausible hypothesis that hot Jupiters are born at large orbital periods, their close-in orbits require a change in orbital energy, or migration, spanning several orders of magnitude. Classic disk-migration theory (\citealt{Goldreich_Tremaine,Lin_Papa}; also cf. \citealt{Julian_and_Toomre_1966}) anticipated that tidal interactions between the protoplanetary disk and planet could lead to inward migration. However, under the assumption that the stellar spin and angular momentum of the protoplanetary disk are aligned, recent observations of spin-orbit misalignment (e.g. \citealt{Winn:2009a,Winn:2010b}) seem to indicate that disk-migration is not responsible for the observed orbits of hot Jupiters.

Tidal dissipation from the star-planet interaction is another potential source of orbital energy loss that leads to inward migration. Given the presence of a distant third body that weakly torques the planet in question, secular changes in orbital angular momentum can lead to close periastron passages---and high eccentricities---where tidal dissipation is activated. Calculations of this process, which we refer to as high-{\it e} migration (HEM), seemed to demonstrate that if the relative strength of tidal dissipation is comparable to the inferred value of the Jupiter-Io interaction \citep{Goldreich_Soter} then a Jovian analogue could migrate from a period of $\sim 5$ yrs to $5$ days in $\sim$ a Gyr \citep{WuMurray, Fabrycky:2007, WuLithwick:2011,Naoz:2011}. Most impressively, spin-orbit misalignment in hot Jupiters was a prediction of HEM \citep{Fabrycky:2007}, and it may be able to account for the formation of all hot Jupiters \citep{Naoz+2013, Li_Naoz+2014}. Also, note that HEM need not always result in spin-orbit misalignment \citep{Petrovich}.

Recently, the plausibility of HEM has been put in question. \cite{Soc12a, Soc12b} have shown that previous calculations of HEM inadvertently over-estimated the strength of tidal dissipation by a few orders of magnitude. Furthermore, \cite{Soc12c} proposed the following test: if HEM is responsible for producing hot Jupiters, there should be a population of super-eccentric migrating Jupiters in the {\it Kepler} sample. However, a statistical analysis of the {\it Kepler} gas giant candidates indicates that such a population is missing \citep{Dawson15}.

In this work, we re-examine the possibility that disk-migration is responsible for forming misaligned hot Jupiters by questioning the assumption that the angular momentum of a protoplanetary disk is initially aligned with that of its host star.
On the basis of observations, it is now accepted that star forming regions are turbulent \citep{MckeeOstriker} and therefore, it is not necessary to assume alignment between stellar spin and angular momentum of the remnant disk. 

Our model of spin-disk misalignment is quite similar to that put forth by \cite{Bate:2010}, although we focus our attention on the outer disk, which contains most of the angular momentum, whereas they restricted their analysis to the inner disk. We find that spin-disk misalignment may be more common than deduced by those authors. We describe the difference between our work and that of \cite{Bate:2010} in \S 4.1. 

The presence of an external body force has been investigated as a means to change the orientation of an initially aligned protoplanetary disk. In particular, \cite{Tremaine1991} suggested that non-axisymmetric collapse of a molecular cloud core could result in spin-disk misalignment. \cite{Lai:2011} raised the possibility that magnetic torques originating from the protostar dictate the disk orientation of the inner-disk, while \cite{BA13} consider gravitational torques resulting from the presence of close-by stellar companion. Additionally, \cite{Terquem} showed that planets that form in a warped disk can be substantially misaligned.

In what follows, we describe hydrodynamic and magnetohydrodynamic simulations that support a turbulence induced spin-disk misalignment scenario consisting of the following ingredients: i) turbulent interstellar material gravitationally collapses toward a common center of gravity, initially forming a disk, ii) as the disk is replenished by the collapsing fluid, its orientation changes, since the converging flow is sourced by a chaotic flow. These simulations produce a distribution of spin-disk misalignments that is consistent with the observed distribution of spin-orbit misalignments. 
Therefore, this scenario may be underlying the observed hot Jupiter spin-orbit misalignments if their migration is due to tidal dissipation in the disk, instead of in the star-planet interaction.

The plan of this paper is as follows: in \S 2, we outline our numerical technique. We describe our results in \S 3, and a discussion of our results and approach, along with a spherical toy model for spin-disk misalignment, are discussed in \S 4, and we conclude with \S 5.

\section{Simulations}
We perform two simulations of star formation in a turbulent molecular cloud clump. The simulations were performed using a grid-based adaptive mesh refinement magnetohydrodynamic code, \textsc{orion} \citep{Klein99, Li2012}. The initial physical properties of the two simulations were identical except for the inclusion of magnetic fields. The non-magnetized simulation will be referred to as the HD run, and the magnetized simulation will be referred to as the MHD run. The finest resolution of the HD run was $\Delta x_{f} = 10 $AU, which corresponds to 5 levels of mesh refinement, and the finest resolution of the MHD run was $\Delta x_{f} = 2.5 $AU, which corresponds to 7 levels of mesh refinement. The finer resolution in the MHD run was needed to avoid the so-called ``magnetic-braking catastrophe'' in which torsional Alfv\'en waves are so effective in transporting away angular momentum that disk formation is prevented \citep{Allen03}. Although the exact mechanism responsible for enabling the formation of protostellar disks in the presence of magnetic fields has not been conclusively proven, the necessity for finer resolution to form disks in magnetohydrodynamic simulations as compared to equivalent hydrodynamic simulations is well established (e.g. \citealp{Seifried12, Myers13}).

To generate turbulent velocity and density initial conditions we adopt the approach used by many groups (e.g. \citealp{Klessen00, Offner09}), in which we divide our simulations into two phases: a \emph{driving} phase, and a \emph{collapse} phase. The driving phase is calculated separately, during which the turbulent initial conditions are generated in the absence of gravity. In the collapse phase the turbulence is no longer driven and self-gravity is enabled, thus leading to star and disk formation.

\subsection{Numerical Methods}
Both the MHD and HD runs were performed in domains with periodic boundary conditions and sides of length $L=81920$ AU = 0.397 pc. They have an isothermal equation of state with temperature of $T=10$ K, and, prior to driving, have a uniform density of $\rho = 1.632\times10^{-19}$ g cm$^{-3}$, which corresponds to a surface density of $\Sigma = 0.2 $ g cm$^{-2}$. The gravitational free-fall time for the box as a whole is $t_{\rm {ff}} \approx 1.65\times10^{5}$ years. The total mass contained in the cloud is 150 $\mathrm{M}_\odot$. The parameters used are consistent with observations of infrared dark cloud cores \citep{ButlerTan} and small massive star forming cores \citep{Mueller02}.

Additionally, the MHD run had an initially uniform magnetic field oriented in the $\hat{z}$ direction with a field strength of $B_\circ = 0.054$ mG. This value was chosen so that the dimensionless mass-to-flux ratio $\mu_\Phi$---the ratio of the mass of a cloud to the maximum mass that can be supported against gravitational collapse by the magnetic field---was 6, which is somewhat larger than the observed norm, ${\langle \mu_\Phi \rangle_{\rm obs} \sim} 2$ \citep{Crutcher2012}, and is large enough that the field does not hinder disk formation (e.g., \citealt{Joos12,ZYLi+2013}).

The driving phase was run using a uniform grid with 512$^3$ cells. The driving pattern was a perturbation cube generated in Fourier space. Power was only injected on large scales---equally balanced between wave numbers satisfying $1\leq {k L}/{2 \pi} \leq 2$. The driving pattern was chosen to have a 2:1 balance of the solenoidal (divergence-free) velocity component to the compressive (curl-free) velocity component \citep{Federrath10}. The turbulence was driven for two crossing times, which allowed the density power spectrum to develop self-consistently. 

The turbulence was driven so that $v_{\rm rms}$ had a sonic Mach number $\mathcal{M} $ of 7.5. This velocity was chosen so that the kinetic energy and gravitational energy in the box as a whole were approximately balanced, which is reflected in the order unity virial parameter \citep{BertoldiMckee92},
\begin{equation}
\alpha_{\rm vir} = \frac{5 \, \sigma_v^2\,  L}{2\, G \,M} = \frac{5 \,\mathcal{M}^2 c_s^2 }{6\, G \,\Sigma\, L} = 1.04.
\end{equation}

When self-gravity was turned on, the base-grid was reduced from $512^3$ to $256^3$ cells and the adaptive mesh refinement was enabled. The reduction of base-grid resolution did not wash out the intricate structures that developed during driving because the mesh was allowed to refine to the first level when neighboring cells had large density, velocity, or magnetic field gradients. The mesh was also allowed to refine down to the finest level (level 5 for HD runs and level 7 for MHD runs) if the density exceeded half of the Truelove-Jeans density on that level. 

\textsc{orion} represents stars as sink particles \citep{Krumholz+04} by modeling their evolution (\citealp{Li2012, Myers13, Lee14}; and refs therein), and, therefore, will be referred to as star particles. Star particles are created when the density exceeded the Truelove-Jeans density on the finest level, where the appropriate Truelove-Jeans density for a magnetized fluid is defined \citep{Myers13}:
\begin{equation}
\rho_{\rm TJ} = \frac{\pi \, c_s^2 J^2}{G (\Delta x_\ell)^2} \left( 1 + \frac{0.74}{\beta}\right).
\end{equation}
Here $\beta$ is the ratio of thermal pressure to magnetic pressure ($\beta=\infty$ in the HD run, so the second term on the right hand side goes away), and $\Delta x_\ell$ is the spatial resolution of the level being considered. Throughout this work we use $J=1/4$. This guarantees that the Jeans length is resolved by at least eight cells for refinement and four cells for sink creation, which has been shown to be sufficient to avoid artificial fragmentation \citep{Truelove+97,Lee14}.

\subsection{Treatment of Angular Momentum}
For the problem at hand we implemented a new way to keep track of the angular momentum accreted onto the star particles throughout the simulations, which represents the only departure from the \textsc{orion} code. Below we briefly describe the treatment of star particle angular momentum. A detailed description of this new method and of the tests performed to verify its accuracy can be found in Appendix A. 
 
The physics governing protostellar angular momentum evolution takes place on a spatial scale smaller (${\lesssim R_\star}$) than is feasible to probe while still following the dynamics of the turbulence in a parsec-scale cloud. Consequently, we use a sub-grid model to approximate the angular momentum transport from the outer edge of the accretion zone to the surface of the star. In reality, on these sub-grid scales angular momentum transport always ensures the specific angular momentum never exceeds the Keplerian value, so we scale the accreted specific angular momentum to the Keplerian value at the surface of the star. The specific angular momentum of a Keplerian orbit increases proportional to $R^{1/2}$. The outer edge of the accretion zone is at $R=4 \Delta x$.
Therefore our sub-grid model reduces the magnitude of the accreted angular momentum by
\begin{equation}
 \frac{l_{\rm out, \, acc} }{ l_{\star \rm \, Kep} }= \sqrt{\frac{4 \Delta x}{R_\star}},
\end{equation}
which is roughly a factor of 23 for the MHD run and 46 for the HD run, assuming a fiducial value of  4 $R_\odot$ for $R_\star$. 

In reality most of the angular momentum that goes into the sink regions should viscously spread back out into the surrounding medium along with a small fraction of the mass. Because our sub-grid model does not account for this it raises the question: how would the results change if less angular momentum was discarded? The amount of angular momentum that is discarded is proportional to $(\Delta x_f )^{1/2}$ where $\Delta x_f $ is the finest resolution of the simulation. Therefore, we were able to address this concern by running a series of three ``zoom-in'' simulations with progressively higher resolution, and demonstrating that our results are sufficiently converged.

\begin{figure}
\begin{centering}
\includegraphics[width=3.5 in]{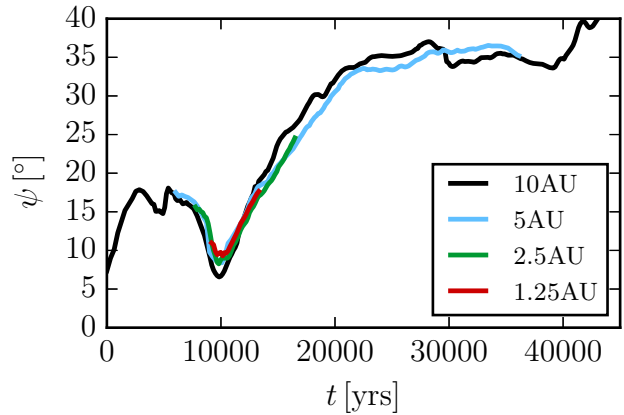}
\caption{The evolution of the stellar obliquity over time for the representative system from the HD run that we focused our zoom-in convergence study on.  The resolution is increased incrementally by a factor of 2,4, and 8 around this star particle and the agreement of the obliquity over time demonstrates that our spin-disk misalignment results are converged. Note this is the same system that is discussed in Section 4.2.\label{converge}}
\end{centering}
\end{figure}

\begin{figure*}
\begin{centering}
\includegraphics[width=\textwidth]{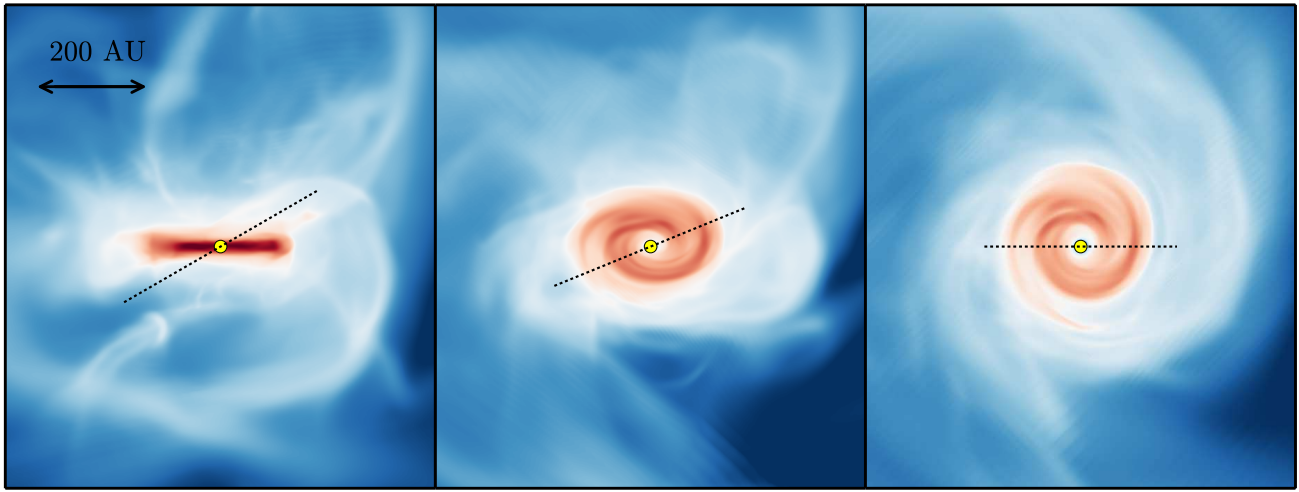}
\caption{An example of one of the disks that formed in the MHD run viewed edge-on (left), at an intermediate angle (center) and face-on (right). Each image is density projection, 800 AU on a side. The dashed line indicates the direction of the stellar spin axis, and its length is the scaled according to the projection of the image. The small yellow dot in the center of each image shows the location of the star particle. The yellow dot is roughly the size of the accretion zone, which for this simulation is 10 AU.\label{Disk_Example}}
\end{centering}
\end{figure*}

The zoom-ins were focused on one star particle from the HD run (the same star particle that is used to demonstrate our toy model in Section \ref{section:toy}) and each had one additional level of mesh refinement. The maximum spatial resolution of the zoom-ins were $\Delta x_f = 5, 2.5,$ and 1.25 AU, which means the amount of discarded angular momentum was decreased by a factor of $2^{1/2}, 2,$ and $8^{1/2}$ respectively relative to the HD run. Of the total angular momentum accreted by the targeted star particle in the 10 AU HD run $\sim80$ per cent entered the accretion zone, $\sim 2$ per cent was accreted onto the star particle and the remaining $\sim 78$ per cent was discarded. We conjecture that the remaining $\sim 20$ per cent was transmitted to the outer part of the disk and transferred to the ambient medium. In the highest resolution zoom-in only $\sim 28$ per cent of the total angular momentum entered the accretion zone, the sub-grid model again ensured $\sim 2$ per cent was accreted onto the star particle while the remaining $\sim 26$ per cent was discarded. In this case, we again conjecture that the remainder of the angular momentum ($\sim 72$ per cent) was transmitted to the outer disk and then to the ambient medium. This conjecture is supported by the agreement of the spin-disk misalignment at all resolutions considered as shown in \ref{converge}.

The first zoom-in was started roughly 6000 years after the star particle formed, the second zoom-in was started 1700 years after the first, and the third was started 1500 years after the second. The temporal spacing was necessary to allow the gas to adjust after the resolution changes. 

To account for the well known non-convergence of simulations of self-gravitating isothermal gas (e.g. \citealt{Kratter10, Krumholz14}), we adopted a barotropic equation of state in the zoom-in simulations. The change in the equation of state prevented disk fragmentation and kept the disks thickness approximately the same. Specifically, we used
\begin{equation}
P = \rho c_s^2 \left[1 + \left(\frac{\rho}{\rho_{\rm ad}} \right)^{2/3} \right], 
\end{equation}
where $\rho_{\rm ad}$ is the critical density, above which the gas becomes adiabatic. We set $\rho_{\rm ad}=2\times10^{-13} {\, \rm g\, cm}^{-3}$. 
We also checked that using this equation of state on the HD run with 10 AU resolution made no difference, which was to be expected because $\rho_{\rm ad}$ is ${\sim 4}$ times larger than the Truelove-Jeans density at this resolution. 

The obliquity of the system is the primary quantity we are interested in. Figure \ref{converge} shows the obliquity over time, and the agreement between the zoom-ins and the HD run is encouraging. The sub-grid model ensures that the star particle accretes the same angular momentum regardless of resolution, so the agreement between the zoom-in simulations indicates that our disks orientations are sufficiently converged. Had we run the highest resolution zoom-ins longer the disparity may have grown, however the fact that all the simulations show the same increase in obliquity at around $10^4$ years after formation demonstrates that roughly the same behavior of a system should be expected regardless of resolution. 

Note that the disk mass and stellar mass do not converge. At higher resolution the disk mass is higher and stellar mass is lower, although the total masses of the systems are the same. This is because by increasing the resolution gas that would have been counted as being in the star is now resolved as being in the disk.

Our sub-grid model also depends on the exact value of the break-up rotation rate of a star, and thus the appropriate magnitude for the cap. We verify that our results do not depend sensitively on the value of the cap we use in Appendix A.

\begin{table}
\caption{Star Formation Properties of the Simulations \label{SF_props}}
\begin{tabular}{cccccc}
\hline
Name & ${t_{\rm f}}/{t_{\rm ff}}$ & $M_{\star,{\rm f}}$ & $N_{\star,{\rm f}}$& $N_{\star, \rm w/ disk}$ & $\epsilon_{\rm ff}$ \\ \hline
MHD      	& 0.87	& 1.22   & 5  	& 2    & 0.03 	  \\ 
HD      	   	& 0.89	& 19.5   & 21  	& 12  & 0.29	  \\ \hline
\end{tabular}
\\{Col. 2: the final simulation time divided by the free-fall time. Col 3.: the total mass in stars at end of the simulation measured in $\mathrm{M}_\odot$. Col. 4: number of stars formed with mass above $0.05$ $ \mathrm{M}_\odot$. Col 5: number of stars with disks. Col 6: The dimensionless star formation rate \citep{KrumholzMcKee05}.}
\end{table}

\section{Results}

In the HD run, 12 of the 21 star particles had disks, which made them suitable for analysis. Those that were not included in the analysis either formed just prior to the simulations end, or had no disk due to fragmentation or stellar encounters in dense regions. In the MHD run 2 of the 5 star particles had disks suitable for study---an example from the MHD run is shown in figure \ref{Disk_Example}. Those omitted either formed too late, or had no disk due to angular momentum loss from magnetic braking. 

This sample of 14 star-disk systems enabled us to determine the effect of accretion from the turbulent interstellar medium (ISM) on spin-disk alignment. Note that with our simulations we did not use the full capability of \textsc{orion} (radiative feedback, protostellar outflows, etc.) as was recently done by \cite{Myers14} since we wished to isolate the effects of turbulent accretion on misalignment. Table \ref{SF_props} lists some of the properties of the simulations. As expected the HD run was more efficient at converting gas into stars than the MHD run.

\begin{figure}
\begin{centering}
\includegraphics[width=3.5 in]{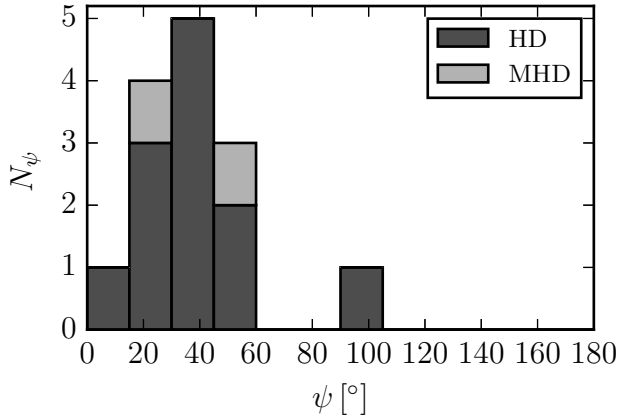}
\caption{The distribution of final spin-disk misalignments as measured by the stellar obliquity $\psi$. The obliquity of the 12 star particles from the HD run are differentiated from the 2 star particles from the MHD run with a darker shade of grey. \label{final_obliq_hist}}
\end{centering}
\end{figure}

\begin{figure}
\includegraphics[width=3.5 in]{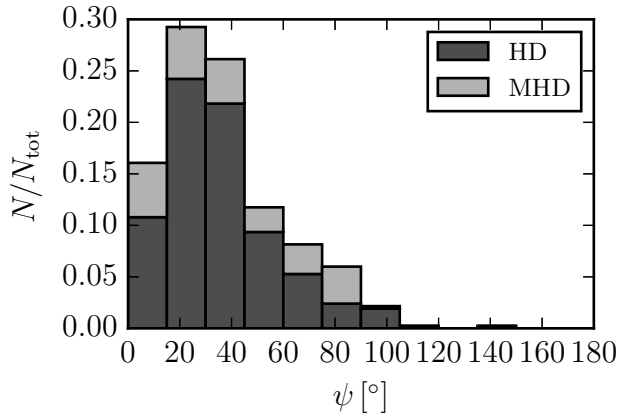}
\caption{The distribution of all spin-disk misalignments as measured by the stellar obliquity. The obliquity of the 12 stars from the HD run and 2 stars from the MHD run are sampled at intervals of $t_d$ if they have a disk. This is to account for the fact that the simulations are ended arbitrarily and accretion has not stopped on its own due to the dispersal of the cloud.\label{Total_Obliq_Hist}}
\end{figure}

\begin{figure}
\begin{centering}
\includegraphics[width=3.5 in]{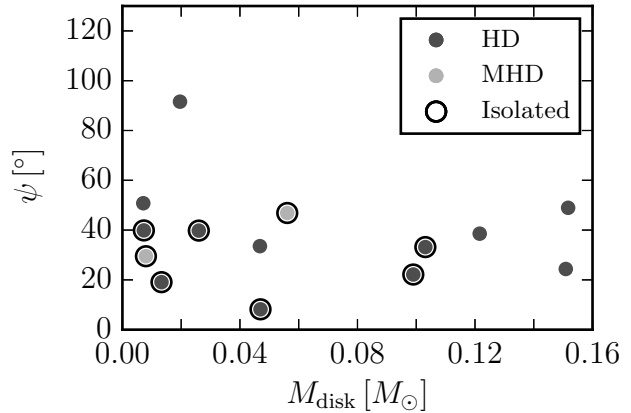}
\caption{The distribution of stellar obliquity relative to disk mass when the simulations were halted. The obliquity of the 12 star particles from the HD run are differentiated from the 2 star particles from the MHD run with a darker shade of grey. The points with a black circle around them correspond to systems that formed in isolation, and never had another star particle within 4000 AU.\label{Oblq_vs_DiskMass_FINAL}}
\end{centering}
\end{figure}

\begin{figure}
\begin{centering}
\includegraphics[width=3.5 in]{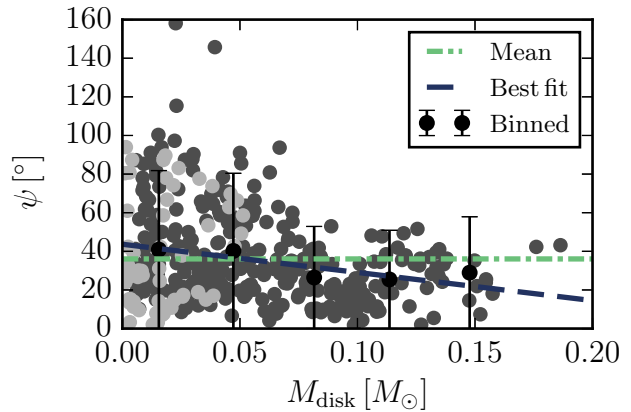}
\caption{The distribution of stellar obliquity relative to disk mass measured at intervals of $t_d$ as in figure \ref{Total_Obliq_Hist}. The light grey points are from the MHD run and the darker grey points are from the HD run. The red points show the mean in values in 5 bins. The error bars indicate the standard deviation within each bin. The blue dashed line shows the best fit linear relation, and the green dash-dot line shows the mean of all values. Both lines are equally consistent with the data. \label{Oblq_vs_DiskMass_ALL}}
\end{centering}
\end{figure}

In our simulations we needed a consistent and rigorous way to identify real, rotationally-supported disks. To do so we adopted a set of five criteria to define the outer boundary of the disk, which are similar to those used by \cite{Joos12}. The criteria are applied separately to concentric rings of material that fall between $R$ and $R + \Delta R$, where $\Delta R$ is set to the resolution of the given simulation. A disk's radius $R_d$ is defined to be the largest radii where all smaller rings meet the criteria. 
The criteria that are applied to the rings are: 
\begin{enumerate}
\item In a Keplerian disk rotational velocity is much larger than the radial velocity, so to be considered part of the disk the ring's average rotational velocity needed to exceed its average radial velocity by a constant factor: $v_\phi > f_{\rm thresh} v_r$. 
\item A Keplerian disk should be close to hydrostatic equilibrium, so we demand that the ring's average rotational velocity exceed the average magnitude of its vertical velocity by a constant factor: $v_\phi > f_{\rm thresh} |v_z|$.
\item To ensure the disks are rotationally supported we require that rotational energy density in a ring exceeds the thermal pressure by a constant factor: $\rho v_\phi^2 /2> f_{\rm thresh} P_{\rm therm}$.
\item The disk should be mostly continuous, so a visual inspection is done to ensure connectivity of the rings, which ensures that distinct parcels of gas are not included in the disk.
\item We also enforce a density criterion of $\rho > 2 \times 10^{-15}$ g cm$^{-3}$ to be consistent with \cite{Joos12}, and which prevents over estimates of the radius by excluding large spiral arms and accretion streams.
\end{enumerate}
We adopt the same constant factor, $f_{\rm thresh} = 2$, for the first three criteria for simplicity, although in principle they could be different.

When star-disk systems accrete varying amounts of angular momentum, first the disk, then the star, adjusts. The rate of change of the stars' and disks' angular momenta vectors are inversely proportional to their masses. Since a star's mass can only increase as the system ages the star will change direction increasingly sluggishly. Alternatively, a disk's mass can fluctuate, so at different times it may be able to rapidly change orientation. In several systems the obliquity goes through multiple cycles of alignment and misalignment as the star catches up with the disk and then falls behind again. Figure \ref{final_obliq_hist} shows the distribution of obliquities of the 14 star particles with disks when the simulations were halted. It is clear that spin-disk misalignment is expected during formation.

\begin{figure}
\begin{centering}
\includegraphics[width=3.5 in]{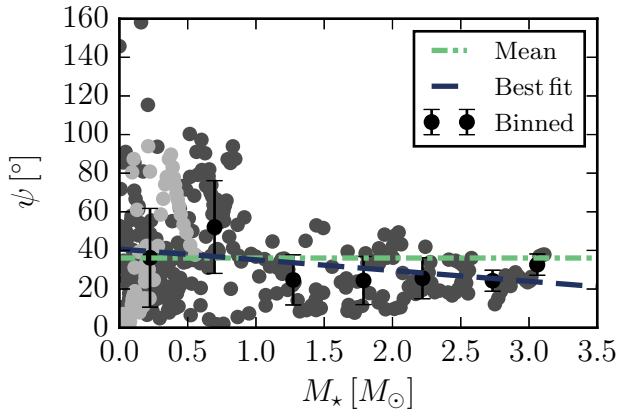}
\caption{The same as fig. \ref{Oblq_vs_DiskMass_ALL} except the obliquity is shown relative to the stellar mass. Again the best-fit line and the mean value are equally compatible with the data. 
\label{Oblq_vs_StarMass_ALL}}
\end{centering}
\end{figure}

The final obliquity in our simulations is not any more physically meaningful than the obliquity at earlier times because the simulations were stopped arbitrarily. In reality a star's obliquity at the end of the embedded phase would be set by when the cloud disperses to the point where accretion onto the system has mostly stopped. Because we did not include any feedback processes accretion would not have stopped until all the gas was in star particles. Therefore, to sample the obliquity distribution in a more representative manner, in figure \ref{Total_Obliq_Hist} we treat measurements of $\psi$ at equally spaced time intervals as distinct. For the time interval we use the average disk accretion timescale---the time for the host star to have accreted as much mass as was contained in the disk---that is defined:
\begin{equation}
t_d = 1500 \left( \frac{ \langle M_d \rangle}{ 0.015 \mathrm{M}_\odot} \right) \left( \frac{10^{-5} \mathrm{M}_\odot \textrm{\ yr}^{-1}}{\langle \dot{M_{\star}} \rangle } \right) \textrm{\ yr}.
\end{equation}
Where the scaling normalizations are the mean values from both of the simulations---$\langle M_d \rangle=0.015 \mathrm{M}_\odot$ and $\langle \dot{M_{\star}} \rangle = 10^{-5} \mathrm{M}_\odot \textrm{\ yr}^{-1}$.
This further demonstrates the general result that spin-disk misalignment is the norm at this phase in stellar evolution. 

Figure \ref{Oblq_vs_DiskMass_FINAL} shows the relationship between star particle obliquity and disk mass in our simulations at the final time, and the points with circles around them formed in isolation, never having another star particle within 4000 AU. This shows that even systems with massive disks and those that had no encounters can be significantly misaligned. Figure \ref{Oblq_vs_DiskMass_ALL} also shows the distribution of obliquity relative to disk mass, but sampled throughout the simulations at intervals of $t_d$, as in figure \ref{Total_Obliq_Hist}. Also shown are the mean obliquity values in 5 bins, as well as the best-fit linear relationship, and the mean value. It is apparent that there is a weak disk mass dependence, although the data are equally consistent with no disk mass dependence. Figure \ref{Oblq_vs_StarMass_ALL} shows distribution of obliquities relative to stellar mass, along with the mean obliquity values in 7 bins, the best-fit linear relationship, and the mean of all values. As with the disk mass there may be a weak dependence of the obliquity on the stellar mass, but the data is equally consistent with there being no dependence. 
Note that because simulations of self-gravitating isothermal gas do not numerically converge in regard to spatial resolution \citep{Krumholz14}, a direct convergence test between star-disk systems in two simulations that differ only in resolution was not possible. Nonetheless, we verified that similar spin-disk misalignment was present in our simulations when we increased or decreased the spatial resolution and when we performed the zoom-in simulations discussed in Section 2.2.

\section{Discussion}

\subsection{Comparison to past work}
\begin{figure*}
\includegraphics[width=\textwidth]{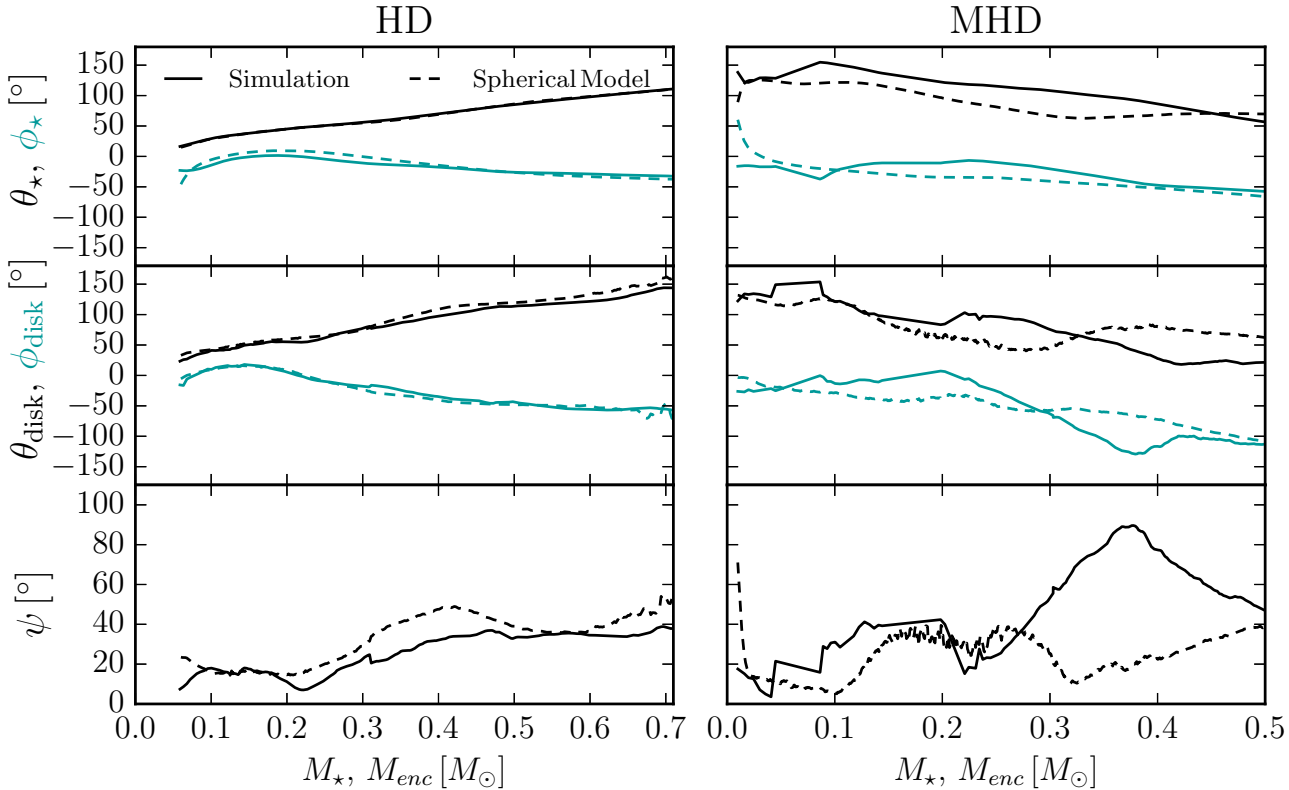}
\caption{The spherical model's predictions for the star particles' (top row), their disks' (middle row) angular momenta direction, and the spin-disk misalignment $\psi$ (bottom row) are compared to the results from the simulations. The left and right column are for the star particle from the \textsc{hd1} and \textsc{mhd1} run respectively. The solid lines show the results from the simulation and evolve as a function of star particle mass. The dashed lines are the spherical model's predictions that change with enclosed mass. The direction of the angular momentum in the top two rows is represented by the polar ($\theta$ in black) and azimuthal ($\phi$ in teal/gray) angle in the grid frame.\label{toy}}
\end{figure*}

Although the physical mechanism we are proposing for spin-disk misalignment is similar to what \cite{Bate:2010} put forth, the approach of our studies differ. We consider the alignment of the outer disk relative to the stellar spin, while \cite{Bate:2010} considered the alignment of inner disk relative to the stellar spin. The difference in our focuses is reflected in our sub-grid models.

To target the inner disk \cite{Bate:2010} used a sub-grid model that viscously evolved the material that was accreted onto their sink particle. Importantly, their model allowed mass and angular momentum to flow outward beyond the sink radius. Unfortunately, it wasn't feasible to introduce material back into the computational domain once it had been removed, so any gas that spread beyond the sink radius in the sub-grid model could not interact with the rest of the simulation. The authors avoided this difficultly as best as possible by limiting their analysis to a single low mass (0.2 $M_\odot$) sink particle that specifically did not have a disk without the post-processing. 
This necessary selection restricted their sample size to a single accretion history that was used for the source term in three variants of their sub-grid model (regular viscosity, high viscosity, and regular viscosity plus disk truncation from stellar encounters). As a result, they did not attempt to predict the distribution of obliquities. Nonetheless, they were able to demonstrate that strong spin-disk misalignment is possible and that it is more likely when the disk is low mass and/or has been disturbed by a stellar flyby. A caveat to their results is that the misaligned disks in their simulations may not be massive enough to form planets; the high viscosity model resulted in $\psi=21.3^{\circ}$ and $M_d = 3 \times 10^{-5} M_\odot = 2.5 \times 10^{-4} M_\star$, the regular viscosity with stellar encounters model resulted in $\psi = 122^\circ$ and $M_d = 2.2 \times 10^{-4} M_\odot = 2.1 \times 10^{-3} M_\star$, while the regular viscosity model without stellar encounters resulted in $\psi = 4.2^\circ$ and $M_d = 3.4 \times 10^{-2} M_\odot = 0.3 M_\star$.

We focus on the outer disks where most of the accretion happens and where most of the angular momentum resides. Our sub-grid model should return similar stellar angular momentum as that of \cite{Bate:2010}, however it does not allow the accreted angular momentum that does not end up on the star to the spread back outside the accretion zone. The benefit of our model is that we are not restricted to studying sink particles that have no disk around them, so we can study systems with large disks and resolve the interaction of the disk with surrounding medium. However, this comes at the cost of no longer conserving angular momentum. The amount of discarded angular momentum decreases with resolution, so the convergence study we discussed in Section 2.2 supports our conviction that our model is sufficiently realistic in the context used here. With our different sub-grid model and larger sample size we find that spin-disk misalignment may be more common than deduced by \cite{Bate:2010}, although our studies are broadly consistent.

\subsection{Simple physical model}\label{section:toy}

To better understand the physical origin of the spin-disk misalignment present in our results, we introduce a simple spherical model, and apply it to our simulation data, thus giving a quantitative demonstration of the connection between turbulence in the ISM and spin-disk misalignment. The fundamental aspect of the physical mechanism is that within a realistic turbulent molecular cloud the angular momentum vector of a protostellar core varies with radius. These variations have been shown to result in time-varying protostellar disk orientations \citep{Smith+11}. We directly link these radial variations in protostellar core angular momenta to changes in star and disk angular momenta, thus demonstrating the connection between turbulence in the ISM and spin-disk misalignment. 

\begin{figure}
\begin{centering}
\includegraphics[width=\linewidth]{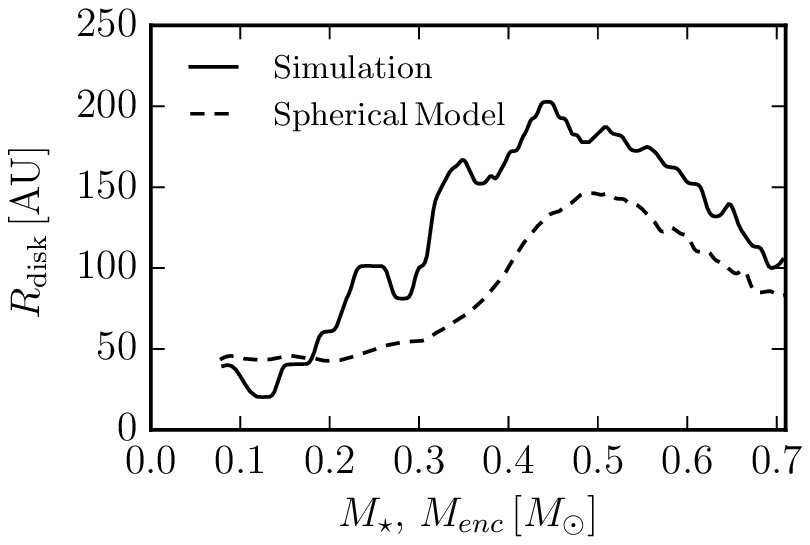}
\newline
\includegraphics[width=\linewidth]{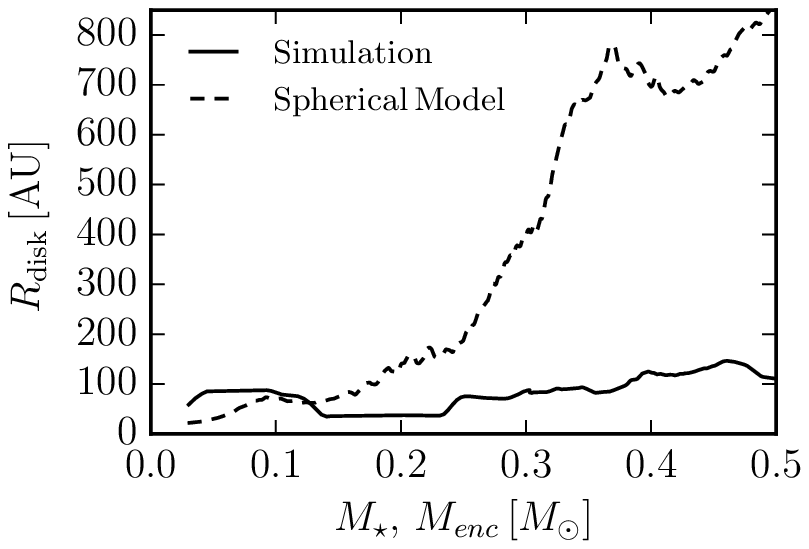}
\caption{The spherical model's prediction for the radius of the disk around the example HD (top) and MHD (bottom). The spherical model does not account for angular momentum transport by magnetic field lines, which accounts for why the prediction in the MHD run is very inaccurate. \label{ToyRadius}}
\end{centering}
\end{figure}

The spherical model we now introduce maps the changes in a core's angular momentum direction with radius---or enclosed mass---to the evolution of the star particle's and its disk's angular momentum. The basic idea of this model is that a star's angular momentum is proportional to the integrated angular momentum of all the gas that it accretes. On the other hand, a disk is constantly gaining and losing angular momentum. Therefore, the angular momentum of the disk is taken to be proportional to that of the recently accreted material. 

In other words, let $\mathbf{L}_d(t)$ be the disk's angular momentum at time some time $t$, and $\mathbf{L}_d(t+t_d)$ be the disk's angular momentum at time $t+t_d$, where $t_d = M_d / \dot{M}_\star$ is the time for the disk to have replaced all of its mass. Then 
\begin{equation}
\mathbf{L}_d(t+t_d) = \mathbf{L}_d(t) + \Delta \mathbf{L}_{\rm acc} + \Delta \mathbf{L}_{\rm ext}, 
\end{equation}
where $\Delta \mathbf{L}_{\rm acc}$ is the angular momentum of the gas accreted in the time $t_d$, and $\Delta \mathbf{L}_{\rm ext}$ is the angular momentum transferred to external matter that time. $\Delta \mathbf{L}_{\rm ext}$ can have contributions from three channels: i) torquing inhomogeneities in the ambient medium, ii) viscous transfer of angular momentum to matter outside the disk, or iii) torques by magnetic fields. We take $\mathbf{L}_d \approx \Delta \mathbf{L}_{\rm acc}$, which is equivalent to the statement that recently accreted angular momentum dominates. This is the essence of how accretion from a turbulent medium leads to changes in disk orientation and is the basis of our spherical model.

With that in mind, and under the assumption that stars form through gravitational collapse (e.g. \citealt{Krumholz+05}), consider a sphere of mass $M_{\rm enc} = M_\star$, with a shell of mass $M_{\rm shell} = M_d$ surrounding it, centered on where a star is about to form. The angular momentum of the star when it reaches a mass $M_\star$ will be proportional to the sphere's angular momentum. Moreover, if the star has a disk around it that has mass $M_d$, then its angular momentum will be proportional to the shell's angular momentum. 
\begin{align}
\mathbf{L}_\star (M_\star) &\propto \mathbf{L}_{\rm sphere}(M_{\rm enc} = M_\star) \\ 
\mathbf{L}_d (M_d) &\propto \mathbf{L}_{\rm shell}(M_{\rm shell} = M_d)
\end{align}

We investigated the ability of this spherical model to predict the angular momentum direction of a star particle and its disk by applying it to a representative star particle from the HD run and the MHD run. Figure \ref{toy} shows our model's prediction for the angular momentum direction of the two exemplar star particles (top panels), for their disks (middle panels), and for the resultant spin-disk misalignment (bottom panels), compared to values obtained in the simulations. By the time these star particles formed the turbulent velocities decayed somewhat. During their formation the Mach number of the whole domain was between $\mathcal{M} = 5.7$ and $5.1$. The model, as simple as it is, does exceedingly well for the HD case and moderately well for the MHD case.

The fact that this simple model can predict the results of the simulations, which followed the small scale details of the gas dynamics, is a testament to the validity of the physical interpretation. 
For the system from the HD run the model's predictions were off by an average of 5$^\circ$, $8^\circ$, and $7^\circ$ for the HD star particle's angular momentum, its disk's angular momentum, and for its obliquity respectively. At the final time these discrepancies were 4$^\circ$, $14^\circ$, and $12^\circ$. 
In the MHD case, the model's predictions was off by an average of 29$^\circ$, $40^\circ$, and $23^\circ$ for the star particle's angular momentum, its disk's angular momentum, and for its obliquity respectively. At the final time these discrepancies were 15$^\circ$, $41^\circ$, and $8^\circ$. 
The model is more accurate in the absence of magnetic fields, which is to be expected because magnetic fields transport angular momentum, which increases the strength of the $\Delta \mathbf{L}_{\rm ext}$ term. The decrease in accuracy when magnetic fields are included does not indicate a diminished accuracy of the physical interpretation. In fact, because magnetic fields link the disk to the more turbulent large scale cloud the effect of turbulence on spin-disk misalignment can be enhanced. 
We can extend this model one step further and use it to predict not only the orientation of the star and disk but also the disk radius. The magnitude of a disk's angular momentum can be written 
\begin{equation}
L_d = k\, M_d\, R_d \, v_{\rm Kep} = k \,M_d \sqrt{G M_\star R_d},
\end{equation}
where $k$ is a dimensionless constant that depends on the structure of the disk and is not known a priori. The disk radius is therefore
\begin{equation}
R_d = \frac{L_d^2}{k^2 M_d^2 G M_\star}.
\end{equation}
Our model assumes that at any given time $\mathbf{L}_d \propto \Delta \mathbf{L}_{\rm acc}$. By taking the stronger assumption that $\mathbf{L}_d \approx \Delta \mathbf{L}_{\rm acc}$ we can predict $R_d$. For the value of $k$ we use the average value of the two disks from each simulation. The disk in the HD run had an average $ k = 0.83$ with a dispersion of 0.03, and the disk in the MHD run had an average $ k = 0.73$ with a dispersion of 0.05. The results of this exercise are shown in figure \ref{ToyRadius} relative to the measured disk radii.

The manner in which the disk radii predictions are wrong are indicative of the relative importance of the other contributions to the disk angular momentum besides what was most recently accreted. Our assumption that $\mathbf{L}_d(t+t_d) \approx \Delta \mathbf{L}_{\rm acc}$ is equivalent to $\Delta \mathbf{L}_{\rm acc} \gg \mathbf{L}_d(t) + \Delta \mathbf{L}_{\rm ext}$. In the HD case, our model only slightly under predicts the disk radius. For the MHD system, our model drastically over predicts the disk radius. This is a result of magnetic braking stripping angular momentum from the disk and allowing the disk to contract, which is beyond the scope of our simple model. 

Our simple model clearly demonstrates the hydrodynamic effect of turbulence on spin-disk misalignment. Moreover, this model indicates that the presence of magnetic fields does not hinder this effect and may in fact enhance it. We have demonstrated that not only is the stellar birth environment's angular momentum imparted to the system as it forms, but that its turbulent motion, which causes variations in the direction of the angular momentum in the natal core, results in the formation of misaligned star-disk systems.

\subsection{Stellar Quadrupole}

A caveat to our results comes from the fact that the young stars we consider are likely to be spinning rapidly, and therefore will have a significant gravitational quadrupole moment. 
The quadrupole of a rotating star will experience a torque from its disk, resulting in the precession of the star. 
This additional coupling of the star and disk can limit spin-disk misalignment if the stellar precession timescale, $t_{{\rm p}\star}$, is much shorter than the disk reorientation timescale, $t_{\rm rd}$ \citep{Lai14,Spaldingea14}. 
Below we present an approximate expression for the stellar precession frequency, and then using a simple post-processing modification to the stellar spin-axis evolution we demonstrate how the obliquity of an example star particle changes given different precession rates.

Unfortunately, the stellar precession timescale is uncertain because it depends strongly on quantities---such as the stellar rotation rate and inner disk radius---that are poorly known for the early stages of protostellar evolution. Although the pertinent physical properties may not be tightly constrained, the interaction between the disk and stellar quadrupole is straightforward. 
We work under the common assumption that the disk is rigid and flat, which is appropriate because of the efficient communication by bending waves and viscous stresses (e.g., \citealt{PapaloizouPringle83,FoucartLai14}) that is also enhanced by self-gravity (e.g., \citealt{TremaineDavis14}). 
The torque from the disk onto the stellar quadrupole is
\citep{Lai14}
\begin{equation}
\mathbf{T} = \int dM_{\rm d} \;\frac{3}{2} \frac{k_q  G M_\star  R_\star^2}{r^3} \frac{\Omega_\star^2}{\Omega_{\rm b}^2} 
\left( \hat{\mathbf{L}}_{\rm d}  \times \hat{\mathbf{L}}_\star \right)
\cos\psi,
\end{equation}
where 
$\Omega_{\rm b}=(GM_*/R_*^3)^{1/2}$ 
is the breakup rotation frequency and $k_q$ is the dimensionless quadrupole moment of the star defined such that $ {I_3 - I_1} = k_q M_\star R_\star^2 (\Omega_\star / \Omega_{\rm b})^2$. Following the lead of \cite{Lai14}, if we assume that the 
surface density of the disk satisfies $\Sigma \propto R^{-1}$ and that $L_\star \ll L_d$, then we find
\begin{equation}
\mathbf{T} = \frac{1}{t_{{\rm p}\star}} \hat{\mathbf{L}}_\mathrm{d} \times \mathbf{L}_\star,
\end{equation}
where 
\begin{equation}
\begin{split}
t_{{\rm p}\star} =  6.0 \times 10^4 \,  & {\rm yr} 
\left( \frac{ P_\star }{3 \mathrm{\, days}} \right)   
\left( \frac{R_\star}{2 R_\odot} \right)^{-1} 
\left( \frac{M_d / M_\star}{0.01} \right)^{-1} \\ \times
&  
\left( \frac{R_{\rm in}}{4 R_\star} \right)^2 
\left( \frac{R_{\rm out}}{100 {\rm AU}} \right)  
\left( \frac{k_\star}{2 k_q} \right)   
\frac{1}{\cos{\psi}},
\end{split}
\end{equation}
and where $R_{\rm in}$ and $R_{\rm out}$ are the inner and outer edges of the disk, respectively, $P_\star$ is the rotation period of the star, and $k_\star$ is the dimensionless moment of inertia of the star defined so that $L_\star = k_\star M_\star R_\star^2 \Omega_\star$.\footnote{This expression for $t_{{\rm p}\star}$ is equivalent to 
(within a factor $2\pi$, which is a matter of definition)
what is derived in \cite{Spaldingea14} in the limit where $\chi/R_{\rm in} \ll 1$, where $\chi = R_\star \left(2 k_q P_{\rm b}/k_\star P_\star\right)^{2/3}$ is the radius of a ring that is inertially equivalent to the rotationally induced bulge on the star.}  For an $n=3/2$ polytrope, $k_q \approx k_\star/2$ \citep{Lai14}.
We have normalized the protostellar radius to $2R_\odot$ based on the results of \citet{Hosokawa+11}, 
who found if accretion is cold then
$R_\star \la 2 R_\odot$ for a wide range of stellar accretion histories and rates.
Hot accretion may cause the protostellar radius to be at most a factor of ${\sim}$2 larger \citep{Hosokawa+11, Baraffe+12}.

We can simplify this expression by assuming that the star co-rotates with the disk at some radius, so that $2 \pi / P_\star = f_\star \sqrt{G M_\star/ R_{\rm in}^3}$. \cite{Long05} found that the co-rotation radius was $\sim1.4 R_{\rm in}$, which implies that $f_\star \approx 0.6$. Thus we have  
\begin{equation}
\begin{split}
t_{{\rm p}\star} = &  3.6  \times 10^4 \,  {\rm yr}
\left( \frac{ P_\star }{3 \mathrm{\, days}} \right)^{7/3}   
\left( \frac{R_\star}{2 R_\odot} \right)^{-3} 
\left( \frac{M_\star}{M_\odot} \right)^{2/3} \\ \times
&  
\left( \frac{f_\star}{0.6} \right)^{4/3} 
\left( \frac{M_d / M_\star}{0.01} \right)^{-1} 
\left( \frac{R_{\rm out}}{100 {\rm AU}} \right)  
\left( \frac{k_\star}{2 k_q} \right)   
\frac{1}{\cos{\psi}}.
\end{split}
\label{eq:tps}
\end{equation}
The normalization for the stellar rotation period is roughly 10 times the breakup rotation period for a solar mass star with $R=2 R_\odot$. 
Observations of the rotation periods of Class II and III protostars
show significant scatter, often displaying a clear bimodality between slow and fast rotators for stars with $M > 0.25 M_\odot$ \citep{Lamm03}. In the Orion Nebula Cluster and in NGC2264 the fast rotators have median period of roughly 3 days \citep{Herbst,Affer13}. 
The rotation periods of Class 0 and I protostars, which are of greater interest to us, are much harder to constrain.  \cite{Covey05} observed a sample of Class I and flat-spectrum protostars spectroscopically and found an
average rotational velocity of 38 km s$^{-1}$
after correcting for inclination; this corresponds to a rotational period of $2.66(R_\star/2R_\odot)$ days. This estimate of the rotational velocity is based on the assumption that these stars can be seen from any angle. In fact, they have accretion disks, which block the equatorial lines of sight with the highest observable rotational velocities, so the actual intrinsic rotational velocity is somewhat larger, and the period somewhat shorter, than this. However, there is no way of knowing how representive this sample is, so this just
provides a specific example of the uncertainties involved in the stellar precession time scale.
Note that if the period is measured directly, Equation (\ref{eq:tps}) shows that the precession time is sensitive to the protostellar radius; however, if the period is measured indirectly by measuring the rotational velocity, then $t_{{\rm p}\star}\propto R_\star^{-2/3}$ is much less sensitive to that uncertain quantity.

\begin{figure}
\begin{centering}
\includegraphics[width=3.5 in]{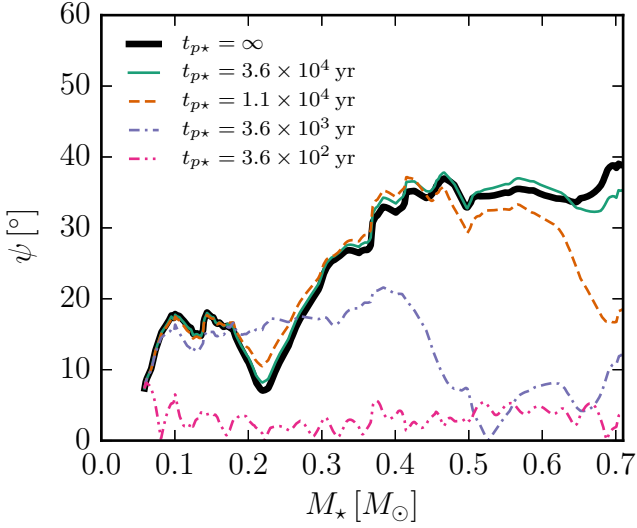}
\caption{Evolution of the spin-disk misalignment angle with different estimates of the stellar precession timescale relative to the star particle's mass. The thick black line with $t_{{\rm p}\star}=\infty$ is the case with no effect from the quadrupole. The solid green line shows the case with a precession timescale equal to our fiducial estimate. The orange dashed, purple dashed-dotted, and magenta dashed-double-dotted lines show the same evolution with a precession timescale $\times 10^{-1/2}, \times 10^{-1},\times 10^{-2}$ our fiducial estimate, respectively. 
\label{quadrupole}}
\end{centering}
\end{figure}

For comparison we measure the disk reorientation time,
\begin{equation}
t_{\rm rd}=\left|\frac{d\hat {\mathbf{L}}_{\rm d}}{dt}\right|^{-1},
\end{equation}
directly in our simulations and find that the average over the growth of all of the stars is $t_{\rm rd} = 1.1 \times 10^{4\pm0.25}$ yr. Although in principle this quantity depends on the disk mass and the accretion rate, in practice variations in $t_{\rm rd}$ are predominately set by the angular momentum accretion history. 

Our fiducial estimate implies that $t_{{\rm p}\star} \approx 3 t_{\rm rd}$; however, it is likely that the timescales may vary significantly, so we cannot claim to have ruled out the possibility that $t_{{\rm p}\star}/t_{\rm rd}$ may be $\lesssim 1$. To address the fact that at this point in a system's evolution the two timescales may be of the same order, and thus not strictly in a limiting regime where one timescale is far shorter than the other, we demonstrate, using a simple post-processing calculation, how the obliquity evolution of an example star particle is modified given different $t_{{\rm p}\star}$ values. In the post-processing we assume that the star's torque on its disk is negligible since $L_\star \ll L_d$. Under this assumption all that changes is the evolution of the stellar angular momentum, which now must include the torque on its quadrupole. The evolution is given by
\begin{equation}
\frac{d {\bf L}_\star}{dt} = \left(\frac{d {\bf L}_\star}{dt}\right)_{\rm acc} + \left(\frac{1}{t_{{\rm p}\star}} \hat{\bf L}_d \times \mathbf{L}_\star\right),
\end{equation}
where the first term on the right hand side is the angular momentum that is accreted onto the star, and the second term is the torque from the disk on the stellar quadrupole.

The system we applied this post-processing to is the same system from the HD run that we discussed in section \ref{section:toy}, which has an average $t_{\rm rd}=1.7 \pm 0.7 \times10^4$ yrs. The evolution of its obliquity is shown in figure \ref{quadrupole} with various values of $t_{{\rm p}\star}$. It is clear from this figure that for our fiducial estimate of $t_{{\rm p}\star}$ the effect of the torque on the star's quadrupole is negligible. Nevertheless, the impact of the quadrupole becomes rapidly important with decreasing $t_{{\rm p}\star}$ to the point that when $t_{{\rm p}\star} \lesssim 10^2$ yr, the system stays nearly aligned throughout its growth. 

Given the uncertainty in protostellar rotation rates, radii, and disk radii, the stellar precession timescale is necessarily very uncertain, and the result of our post-processing calculation demonstrates the importance of better constraining these values both observationally and theoretically. 
\cite{Spaldingea14} 
adopted parameters quite different from ours: They considered
the maximally precessing case in which the star rotates at the breakup rotation rate and the inner edge of the disk is in contact with the star; they also adopted a characteristic protostellar radius of $4R_\odot$. For these conditions, we too find that 
$t_{{\rm p}\star}\ll t_{\rm rd}$ so that the stellar spin remains closely aligned with the rotation axis of the disk.  We have presented evidence for the values we have adopted above, which lead to the opposite conclusion. Theoretically, \cite{MKLin_ea11} have argued that gravitational torques are sufficient to ensure that a protostar never rotates at more than 1/2 of its breakup rotation rate, and \cite{Shu1994} have shown that stellar magnetic fields $\sim 1$kG (contrary to the MG fields claimed by \cite{Spaldingea14}) are sufficient to truncate the disk before it comes into contact with the star for protostellar accretion rates 
$\la 10^{-5} M_\odot$~yr$^{-1}$. 

\subsection{Comparison to Observations}

To date no observations have been made of the stellar spin axis in systems as young as those in our simulations, so a direct comparison of our spin-disk misalignment results is not possible. However, the stellar obliquity has been measured relative to planetary orbits in more evolved systems in which the disk has dissipated. Therefore, a comparison between our spin-disk and the observed spin-orbit misalignments may indicate to what extent the inclination of the planetary orbits is inherited from the disks that they formed in. This can then be used to discern the dominant channel of hot Jupiter migration because in the HEM paradigm the misalignment is due to post planet formation evolution whereas in the disk-migration paradigm a planet's inclination is the same as that of the disk.

To compare our results for the distribution of stellar obliquities $\psi$ to what has been observed we must take into consideration that it is only possible to measure the plane-of-sky projected spin-orbit misalignment angle $\lambda$ (note that some authors use $\beta$ instead). 
The relation between $\psi$ and $\lambda$ is simplified by the fact that the methods for measuring $\lambda$ require the planet to transit the star so we can assume that our disks' angular momenta are in the plane of the sky. \cite{FabryckyWinn} showed that in this case the probability distribution function of $\lambda$ for a given $\psi$ is
\begin{equation}
P(\lambda|\psi) = \begin{cases}\displaystyle
 \frac{2}{\pi} \frac{\cos \psi}{\cos \lambda (\cos^2 \lambda - \cos^2 \psi)^{1/2}}, \\& \hspace{-1.2cm}|\lambda - \frac{\pi}{2}| \geq |\psi - \frac{\pi}{2}|,\\
 \qquad \qquad0, \\&\hspace{-1.2cm} |\lambda - \frac{\pi}{2}| < |\psi - \frac{\pi}{2}|.
\end{cases}
\end{equation}

\begin{figure}
\begin{centering}
\includegraphics[width=3.5 in]{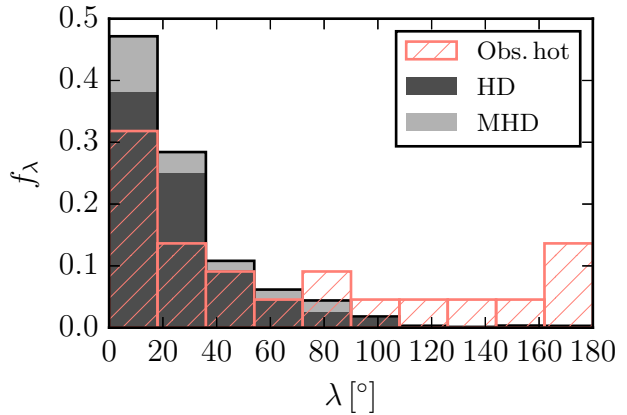}
\caption{The distribution of observed projected spin-orbit misalignment angle $\lambda$ for hot stars ($T_{\rm eff} \gtrsim 6250$ K) as listed on \url{exoplanets.org} as of March 2015 is shown in magenta. The stacked gray histogram shows the predicted distribution of $\lambda$ derived from the combination of our simulations. The dark and light grey portions reflect the fraction in each bin from the HD and MHD runs respectively. \label{comparison to observations}}
\end{centering}
\end{figure}

Approximately one-third of the 68 measurements of $\lambda$ in exoplanet systems listed on \url{exoplanets.org} have substantial misalignment with $|\lambda| > 30^\circ$, and ${\sim14}$ per cent are retrograde with $|\lambda| > 90^\circ$. After applying equation 17 to our simulations, they predict $P(\lambda>30^\circ) = 34$ per cent, in good agreement with the observations. Our simulations also predict $P(\lambda>90^\circ) = 3$ per cent, which is significantly less than the observed value. For this comparison we have combined the samples from our two simulations, which are sufficiently similar, into a single sample to improve the statistics. 
However this may not be the most salient comparison to make. It is well established that hot Jupiters around ``hot'' stars ($T_{\rm eff} \gtrsim 6250$ K) are much more likely to be misaligned than those around cooler stars, which are almost always aligned ($\lambda <30^\circ$) \citep{Schlaufman2010,Winn:2010b}. A possible explanation for this dichotomy is that dissipation in the stellar convective envelope of the star--planet tidal interaction, which is more effective in cooler stars that have larger convective envelopes, has had enough time to realign the planets around cool stars (e.g., \citealt{Albrecht:2012}). In that case, our simulations are most relatable to the systems which have not undergone significant dissipation---namely, the hot stars. Figure \ref{comparison to observations} shows the projected spin-orbit misalignment distribution from our simulations relative to the observed values from systems with stellar $T_{\rm eff} \geq 6250$ K.

Our simulations were intended to be a proof of concept for turbulence induced spin-orbit misalignment, so the level of agreement between our results and observations are only to be taken as rough indicator. Moreover, this mechanism sets the initial conditions for all future influences on the misalignment. One such influence that may alter a system's spin-orbit misalignment is the angular momentum transport by internal gravity waves in massive (hot) stars with convective cores and radiative envelopes \citep{RogersLin2012, RogersLin2013b}. 
Additionally, interactions with a stellar magnetic field \citep{Lai:2011}, or a binary companion star can drastically alter the spin-orbit misalignment \citep{BA13, Lai14, CridaBatygin}, although observations do not show a correlation between the presence of a directly imaged companion and spin-orbit misalignment \citep{Ngo2015}.
Therefore by taking the obliquity distribution from our simulations as initial conditions---instead of assuming all systems begin aligned---the continued evolution from effects such as these may introduce the retrograde tail missing from our distribution as well as possibly improving the fit between the predicted $\lambda$ distribution from these mechanisms and observations.

Although there are many complicating factors in this very active field of research, the comparison between our simple model and the observations indicates that 
a significant contribution to
the observed spin-orbit misalignments could originate in the disk when it forms. 
This supports the possibility that the migration of misaligned hot Jupiters could 
be due to tidal dissipation in the disk, rather than tidal dissipation in the star-planet interaction.
And, putting aside the question of migration, our results demonstrate that the process of star formation can lead to significant spin-orbit misalignment by the time planets form, thereby setting the initial conditions for subsequent evolution of the misalignment.  

\section{Conclusions}

Star formation in a turbulent environment may lead to spin-orbit misalignment. We show this by performing grid-based simulations of star formation in turbulent molecular clouds with realistic initial conditions. The stellar angular momentum is approximately parallel to the total angular momentum of the material it accreted, whereas the protoplanetary disk's angular momentum at a given time is much more variable and is determined by the most recently accreted matter. A simple spherical accretion model is remarkably successful in predicting the orientation and radius of the disk in the absence of magnetic fields. 
Additionally, we verify that including the effect of the stellar quadrupole does not wipe out the misalignment at this stage as long as the protostar rotates considerably slower than the breakup rotation rate. 

The results of our simulations are consistent with the idea that the process of star
formation in a turbulent medium could set the initial conditions for
spin-orbit misalignment. 
This concept of a non-zero initial obliquity, especially if taken in concert with other mechanisms that can further modify spin-disk misalignment, may explain the observed spin-orbit misalignment distribution, particularly for systems with ``hot'' stars ($T_{\rm eff, \star} \geq 6250$ K) that have likely not undergone significant dissipation and realignment. 
This supports the possibility that a substantial fraction of misaligned hot Jupiters have undergone disk-migration in a misaligned disk as opposed to HEM, and that the initial conditions for the evolution of spin-disk or spin-orbit misalignment should come from a range of misalignments. 

Interaction between the accretion disk and the stellar quadrupole created by rapid rotation of the protostar will cause the stellar spin axis to adiabatically trail the disk axis if the precession period is significantly less than the disk reorientation time scale \citep{Spaldingea14}. The precession period depends on several quantities that are not well-determined observationally: the rotational period of the star, the inner radius of the disk, and the radius of the protostar. Since planets form at the end of the accretion phase, these quantities must be determined over the last disk reorientation time, when the protostar is accreting its final tens of percent of its mass---i.e., most likely during the Class~I stage. We presented evidence that the rotational period of protostars in this phase is slow enough for our mechanism to work, but the sample is not unbiased and more observations are needed to confirm this.

Our results demonstrate a mechanism for spin-orbit misalignment in hot Jupiters {\it in isolation} i.e., in the absence of a third body. There are many questions that remain. How are the statistics of the turbulent flow of the birth cloud reflected in the angular momentum of the stellar host and planetary orbits? How does the stellar quadrupole affect spin-disk misalignment? To what extent do interstellar magnetic fields affect spin-disk misalignment by communicating angular momentum between the protoplanetary disks and turbulent ISM? Why are the orbits of many hot Jupiters aligned? Is spin-orbit alignment more prevalent in massive stars or low-mass stars \citep{Dawson14}? What shuts migration off so that the hot Jupiter does not plunge into the stellar host? 

\section*{Acknowledgements}
We wish to thank Fred Adams, Konstantin Batygin, Daniel Fabrycky, Eve J. Lee, Yoram Lithwick,  
Kevin Schlaufman, and Scott Tremaine for useful discussions regarding the underlying physical implications of this work while this paper was in preparation. C.F.M. would like to thank Scott Tremaine and the Institute for Advanced Study, where this project began, for their hospitality. D.B.F. wishes to thank William J. Gray, Pak-Shing Li and Andrew Myers for their computational expertise, and helping run and analyze the simulations presented in this paper. Support for this work was provided by the National Science Foundation through grant AST-1211729 (C.F.M. and R.I.K.); by NASA through grant ATP-
NNX13AB84G (R.I.K., C.F.M.); 
and by the US Department of Energy at the Lawrence Livermore National Laboratory under contract DE-AC52-07NA27344 (A.J.C., R.I.K.). 
Additionally, D.B.F. is supported by the Berkeley Fellowship for Graduate Study and the National Science Foundation Graduate Research Fellowship under grant number DGE-1106400. 
This research was supported by a grant of high performance computing resources from the National Center of Supercomputing Application through grant TG-MCA00N020, under the Extreme Science and Engineering Discovery Environment (XSEDE), which is supported by National Science Foundation grant number OCI-1053575.
This research has made use of the Exoplanet Orbit Database and the Exoplanet Data Explorer at exoplanets.org \citep{exoplanets.org}.
We have also made extensive use of the yt toolkit \citep{TurkYT} for data analysis and plotting.

\bibliographystyle{mn2e}
\bibliography{Fieldingea14}

\begin{appendix}
\section{Star Particle Angular Momentum}
Star particles in \textsc{orion} can accrete angular momentum in two ways: through gas accretion, and by merging with another star particle. \cite{Krumholz+04}, \cite{AJC12}, and \cite{Lee14} explain the algorithms used in \textsc{orion} for gas accretion and particle mergers in great detail. For the present study we ensure that the angular momentum accretion is invariant under Galilean transformations, and that the amount of angular momentum accreted did not exceed simple physically reasonable limits. Here we describe the newly implemented star particle angular momentum method followed by a description of the tests performed to ensure its accuracy and that our results are insensitive to the cap used to limit the magnitude of accreted angular momentum.

We start with a description of gas accretion. For our purposes it suffices to say that gas accretion proceeds by removing a fraction of the matter in each cell falling within the accretion zone around each star particle and adding it to the star particle. The accretion zone is defined to be all cells within $r_{\rm acc} = 4 \Delta x$, where $\Delta x$ is the grid size on the finest level. 
Once the amount of mass to be accreted by a star particle from each cell, $m_{\rm accr,cell}$, has been determined, the angular momentum of the matter is calculated relative to the star particle:
\begin{equation}
\mathbf{L}_{\rm acc} = \displaystyle\sum\limits_{\rm acc,cell} m_{\rm acc,cell} \big[ ( \mathbf{r}_{\rm cell} - \mathbf{r}_\star) \boldsymbol{\times} ( \mathbf{v}_{\rm cell}- \mathbf{v}_\star ) \big].
\end{equation}
However, this is not what gets added to the star particle. Material orbiting a star particle at the outer edge of the accretion zone has a much higher specific angular momentum than the maximum breakup specific angular momentum of the star. To avoid from having our star particles rotating faster than breakup, we use the sub-grid protostellar model \citep{Krumholz+04}, which evolves---among other quantities---the stellar radius $R_\star$, to set an upper limit on the magnitude of the accreted angular momentum. We set the maximum accreted angular momentum to be the angular momentum the same amount of mass would have if it were orbiting at the Keplerian velocity at the surface of the star. Therefore the angular momentum that is accreted to the star is:
\begin{equation}
\Delta\mathbf{L}_\star = \textrm{min} \left\{ \left|\mathbf{L}_{\rm acc}\right|, M_{\rm acc} R_\star \sqrt{\frac{G M_\star}{R_\star}} \right\} \frac{\mathbf{L}_{\rm acc}}{\left|\mathbf{L}_{\rm acc}\right|}
\end{equation}
This capping procedure is an upper limit that approximates the intricate details of the actual angular momentum transport that take place on these scales. In the context of the work presented in this paper, the star particle accretes the most angular momentum that is physically reasonable.

Star particles are merged when two star particles come within $r_{\rm merge} = 8 \Delta x$ of each other provided the mass of the less massive particle is no greater than $M_{\rm merge,\, max} = 0.05 \mathrm{M}_\odot$. The angular momentum of the merger product is a combination of the internal angular momenta of the star particles and their mutual orbital angular momentum. The orbital contribution is given by
\begin{equation}
\mathbf{L}_{{\rm orbit} } = \mu \left(\mathbf{r}_{1}-\mathbf{r}_{2}\right) \boldsymbol{\times} \left(\mathbf{v}_{1}-\mathbf{v}_{2}\right),
\end{equation}
where $\mu$ is the reduced mass. 

As in gas accretion case, the angular momentum of the star particle is capped. The newly merged star particle's angular momentum before the cap is
\begin{equation}
\mathbf{L}_{{\rm precap} } = \mathbf{L}_{{\rm orbit} } + \mathbf{L}_{1}+\mathbf{L}_{2 },
\end{equation}
where $\mathbf{L}_{1}$ and $\mathbf{L}_{2 }$ are the internal angular momenta of the star particles before the merger. If this angular momentum exceeds the breakup rate then it is scaled down, so that
\begin{equation}
\mathbf{L}_{\rm merge} = \min \left\{\left|\mathbf{L}_{\rm precap}\right|,L_{\rm breakup} \right\} \frac{\mathbf{L}_{\rm  precap}}{\left|\mathbf{L}_{\rm precap}\right|}.
\end{equation}
In our model, we use $L_{\rm breakup} = \sqrt{ G M_\star^3 R_\star}$.

\subsection{Test Problems}
To verify that the above star particle angular momentum implementation performs as desired we developed two tests. The tests were designed to individually test the gas accretion and merger angular momentum algorithms. Specifically, we ensured that the direction and magnitude of the star particle angular momentum matched analytic results, that the results are completely invariant under Galilean transformations, and that the results did not depend on resolution. Additionally, we verified that altering the angular momentum cap significantly does not alter our results substantially, nor change our conclusions. Below we describe the two test problems, the various conditions under which they were executed, and the results of the tests. 

The gas accretion angular momentum test consisted of a box with uniform density of $\rho = 10^{-11}$ g cm$^{-3}$ rotating with constant angular momentum about a non-grid axis with a star particle in the center that has an initial mass $M=0.5 \mathrm{M}_\odot$. We also tested the case where the rotation was about the $z$ axis. Additionally, in some tests we gave both the gas and the star particle a velocity---subsonic and supersonic---in an arbitrary direction to ensure that the results were not changed when center of mass was moving. And, of course, we varied the resolution, using a base grid of $32^3$, up to $256^3$. In all cases we knew the desired magnitude and direction of the specific angular momentum of the gas, which allowed us to determine the accuracy of the specific angular momentum accreted onto the star particle. For every test we ran, the direction of the angular momentum differed from the predicted direction by less than $10^{-4}$ degrees, and the magnitude differed by at most $\sim4.5 $ per cent. The magnitude was always less the predicted value, but came into $<1$ per cent agreement as the resolution was increased and as the velocity of the imposed bulk motion was dropped. This is indicative of the fact that the small inaccuracy in the magnitude of the angular momentum is due to the inherent numerical errors, which cannot be avoided but that become increasingly unimportant when using production run levels of resolution.

The setup for the merger accretion test problem is as follows: two star particles are placed in orbit within the merging length scale so that they merge immediately, and the pair is given a net velocity in an arbitrary non-grid-axis direction (both sub- and supersonic velocities were tested). The normal vector of the orbit of the two stars points in a different non-grid-axis direction. To test if the angular momentum cap was working we varied the star particles' radii. With physically reasonable radii the cap is triggered, and with extremely large radii the cap is not triggered so the sum of angular momentum vectors should be returned. In every combination of star particle radii, net velocity, resolution, mass ratio, and orbital angular momentum direction, the direction and magnitude of the resultant angular momentum agreed with the predicted value to machine accuracy. 

An additional, and important, test we performed was the sensitivity of our results on the value used to cap the magnitude of accreted angular momentum. The cap we used was the simplest physically reasonable choice, and was equivalent to our star particles having a moment of inertia $I_\star = M_\star R_\star^2$. This original cap ensured the angular momentum of a star particle was always below the breakup limit of $L_{\rm breakup, \, orig} = \sqrt{G M_\star^3 R_\star}$. Because of the crudeness of the approximation we wanted to make sure that our results did not change if we used a different value. To do so we re-ran a portion of both of our simulations using two alternative caps. One alternative cap had no limit at all, so its breakup rate was infinite. The other cap was 5 times smaller than what we used in our production runs. This second cap corresponds to the star particles having $I_\star = 0.2 M_\star R_\star^2$, which is approximately what one would expect for an $n=3/2$ polytrope, and therefore applies to fully convective proto- and pre-main-sequence stars \citep{BA13}. With this cap the star particles have a breakup magnitude of $L_{\rm breakup, \, poly} = 0.2 \sqrt{G M_\star^3 R_\star}$.

Using both of these alternative angular momentum caps we ran the HD simulation long enough for the most massive star to reach nearly half a solar mass. In this time seven star particles formed, five of which were above 0.05 $\mathrm{M}_\odot$. The change in cap, as expected, changed the magnitude of star particles' angular momenta. However, it is reassuring that the directions were not affected substantially. The angular difference between most massive star particle's angular momentum vector using the original cap and polytrope cap was on average 0.3$^\circ$ with a maximum of $0.5^\circ$. For the same star particle, when comparing the original cap to using no cap the average difference rose to $2.8^\circ$, with a maximum of $4.0^\circ$. The average angular discrepancy for all five star particles with masses above 0.05 $\mathrm{M}_\odot$ when using the original cap and polytrope cap was $3.4^\circ$, and when using the original cap and no cap the average was $4.2^\circ$. 

Likewise, we re-ran the MHD simulation long enough for the most massive star to reach more than half of its final mass. In that time only one star particle formed. The average angular difference between the star particle's angular momentum vector using the original cap and polytrope cap was $0.5^\circ$ with a maximum of $1.1^\circ$. The average and maximum angular difference when using the original cap and no cap were $2.2^\circ$ and $5.2^\circ$, respectively. It is clear from these tests that our results and conclusions would not change significantly if we used a different value for the cap, or even no cap at all.

\end{appendix}
\end{document}